\documentclass{article}

\usepackage{PRIMEarxiv}

\usepackage[utf8]{inputenc} 
\usepackage[T1]{fontenc}    
\usepackage{hyperref}       
\usepackage{url}            
\usepackage{booktabs}       
\usepackage{amsfonts}       
\usepackage{nicefrac}       
\usepackage{microtype}      
\usepackage{lipsum}
\usepackage{fancyhdr}       
\usepackage{graphicx}       
\graphicspath{{media/}}     
\usepackage[table,xcdraw]{xcolor}
\usepackage{tcolorbox}
\usepackage{titlesec}
\title{Privacy-Preserving Methods for Bug Severity Prediction
}

\author{
  Havvanur Dervişoğlu, Ruşen Halepmollası \\
  Istanbul Technical University \\
  İstanbul, Türkiye \\
  \texttt{\{dervisoglu25, halepmollasi\}@itu.edu.tr} \\
   \And
  Elif Eyvaz \\
  Scientific and Technological Research Council of Turkey - TÜBİTAK \\
  Kocaeli, Türkiye\\
  \texttt{elif.eyvaz@tubitak.gov.tr} \\
}

\begin{document}
\maketitle

\begin{abstract}

Bug severity prediction is a critical task in software engineering as it enables more efficient resource allocation and prioritization in software maintenance. While AI-based analyses and models significantly require access to extensive datasets, industrial applications face challenges due to data-sharing constraints and the limited availability of labeled data.  In this study, we investigate method-level bug severity prediction using source code metrics and Large Language Models (LLMs) with two widely used datasets. We compare the performance of models trained using centralized learning, federated learning, and synthetic data generation. Our experimental results, obtained using two widely recognized software defect datasets, indicate that models trained with federated learning and synthetic data achieve comparable results to centrally trained models without data sharing. Our finding highlights the potential of privacy-preserving approaches such as federated learning and synthetic data generation to enable effective bug severity prediction in industrial context where data sharing is a major challenge.

 The source code and dataset are available at our GitHub repository: \url{https://github.com/drvshavva/EASE2025-Privacy-Preserving-Methods-for-Bug-Severity-Prediction}.

\end{abstract}


\section{INTRODUCTION}
Bug severity prediction plays a crucial role in software development and maintenance \cite{gomes2019bug}. It assists teams in identifying and addressing critical issues more effectively. Moreover, the success of prediction models largely depends on access to large-scale datasets. 

While the availability of open source projects has enabled the widespread use of artifacts from these projects in software engineering research, advancements have been constrained by unavailability of data from non-open source repositories \cite{shanbhag2022exploring}. A major cause of limited data availability in industrial environments is often privacy concerns and data-sharing restrictions.

To address these challenges, we explore the use of privacy-preserving methods such as Federated Learning (FL) and synthetic data generation for bug severity prediction. FL is a distributed Machine Learning (ML) algorithm that enables collaborative model training across multiple clients without sharing data \cite{yang2019federated}. FL, which can be applied to almost every ML-based application to ensure data privacy \cite{narula2024comprehensive}, is widely utilized in various domains such as software, finance, health, and IoT \cite{lo2021systematic, aktas2024, ulver2023federated}. Furthermore, synthetic data generation is another innovative approach that addresses data privacy and security issues \cite{razi2025enhancing}. Synthetic data generation refers to data obtained through a generative process that captures the statistical properties and patterns of real data while ensuring the absence of real sensitive information \cite{assefa2020generating}. The use of synthetic data can mitigate the challenge of limited data availability and can also be utilized to mitigate bias, augment real datasets, and increase model robustness \cite{jordon2022synthetic}. Both FL and synthetic data generation provide results comparable to centralized learning while adhering to privacy regulations and mitigating data access constraints.

In this study, we compare FL and synthetic data generation approaches against centralized learning for method-level bug severity prediction by employing ML models. To ensure a robust evaluation, we build upon the study in \cite{mashhadi2023method} that centrally employed  ML models. Our research extended this by considering industrial cases where data privacy and accessibility constraints are main issues. In this context, we define the following RQ:
\begin{itemize}
    \item Can privacy-preserving methods achieve results comparable to centralized training to predict bug severity without sharing data?
\end{itemize}

By addressing this RQ, we aim to explain the effectiveness of privacy-preserving methods, such as FL and synthetic data generation, to achieve performance comparable to centralized training for bug severity prediction without data sharing. We utilized the same benchmark datasets in the prior study, namely Defects4J and Bugs.jar. We evaluate various preprocessing stages, including scaling, feature selection and dimensionality reduction methods, and classification algorithms to obtain the most effective pipeline. Our results, measured using F1-score, Matthews Correlation Coefficient (MCC), Cohen's Kappa Score, and Geometric Mean (G-Mean), demonstrate that FL and synthetic data generation achieve comparable results to centralized learning.

The remainder of this paper is structured as follows: Section 2 provides background and reviews related work, Section 3 presents the research design, Section 4 details the datasets and experimental setup, Section 5 discusses the findings, and Section 6 concludes with key findings and future research directions.

\section{RELATED WORK}
Bug reports are commonly categorized by severity levels, from critical to minor. Critical bugs represent the most serious issues, whereas trivial bugs result in minimal user disruption. To prioritize bug reports effectively, researchers have developed various ML, Deep Learning(DL), and LLMs based approaches. 

For instance, Kim and Yang\cite{9878105}
classified bug reports by topic in the Eclipse and Mozilla open-source projects to prevent the misclassification of bug severity due to subjective assessments. They applied feature selection for each topic and predicted severity using a CNN-LSTM model. Their approach outperformed baseline models such as DeepSeverity and EWD-Multinomial\cite{9878105}. Asif et al.\cite{ali2024bert} introduced BERT-SBR, a fine-tuned BERT-based approach that incorporates sentiment analysis and word embeddings to improve bug severity prediction results. According to their experimental results, BERT-SBR significantly outperformed existing DL classifiers\cite{ali2024bert}. 

Moreover, in our baseline study, Mashhadi et al.\cite{mashhadi2023method} investigated the use of source code metrics, source code representations with LLMs, and their combination for bug severity prediction. They conducted experiments on the Defects4J and Bugs.jar datasets. Their findings show that fine-tuning CodeBERT significantly improved the prediction performance compared to traditional ML models.

FL has emerged as a significant alternative in industrial software development environments, as it enables model training without data sharing and preserves data privacy. For this purpose, Yang et al. \cite{yang2024federated} analyzed the impact of FL on code clone detection and defect prediction using the CodeXGLUE and PROMISE datasets. Authors compared centralized learning and FL approaches and demonstrated that FL can facilitate bug prediction without requiring data sharing. 

Lin et al.\cite{lin2024open} discussed how FL can be used to collaboratively develop and maintain open-source AI code models while preserving data privacy and security. They extensively analyzed the impact of diverse data sources, label imbalance, and data heterogeneity on FL model performance and accuracy.
Similarly, Hu et al.\cite{hu2025fedolf} aimed to improve FL model accuracy under imbalanced data distributions. The study introduced the TabDiT data balancing method and a parameter aggregation strategy based on information entropy. They demonstrated that FL can achieve accuracy scores comparable to centralized learning without requiring data sharing.

Synthetic data generation plays a crucial role in ML applications by addressing various issues, such as data scarcity, class imbalance, and privacy constraints \cite{jordon2022synthetic}. Soltana et al.\cite{8115698} proposed an approach for generating synthetic test data that is statistically representative and logically valid for data-intensive systems. They found that their approach is scalable and capable of simultaneously fulfilling both statistical representativeness and logical validity requirements\cite{8115698}.

Although bug severity prediction has been extensively studied by researchers, there is a gap in the literature when it comes to comparing centralized methods with privacy-preserving approaches, such as FL and synthetic data. This study aims to address this gap by investigating and comparing these methods in the context of bug severity prediction.

\section{METHODOLOGY}
In this section, we present the dataset, preprocessing steps, feature extraction methods, classification algorithms, and training strategies used in this study.

\subsection{Dataset}
We utilized publicly available datasets, namely Defects4J\cite{just2014defects4j} and Bugs.jar\cite{saha2018bugs}. In this study, when predicting bug severities, we used a merged version of these datasets\cite{mashhadi2023method}. The final dataset consists of 3,342 instances labeled across four severity levels (0 being the most critical and 3 the least critical), with a class distribution of 275, 2082, 291, and 694 instances, respectively. 

\subsection{Preprocessing and Feature Selection}
We applied software metrics for ML model training and source code for CodeBERT fine-tuning. We evaluated the effects of different methods on model performance to determine the best preprocessing and feature selection approaches for software metrics. When applying preprocessing steps, we used the eXtreme Gradient Boosting (XGBoost) classifier with default parameters and tested various scaling methods, including StandardScaler, MinMaxScaler, MaxAbsScaler, and RobustScaler. For feature selection, we applied SelectKBest, Random Forest (RF), recursive feature elimination, variane threshold, L1-based selection, and minimum redundancy maximum relevance. Then, we evaluated the impact of feature reduction and polynomial features based on classification results. We compared the model outputs and achieved the best results using RobustScaler for scaling and SelectKBest for feature selection. Also, polynomial features provided the best performance, with a polynomial degree set to 3 and selector\_k to 9 through hyperparameter optimization.

For the CodeBERT model, we only used source code texts as input. During preprocessing, we tokenized the code texts using the RoBERTa-based CodeBERT tokenizer and applied padding to complete the 512 block size. In the feature extraction stage, we utilized the CodeBERT model to generate context-aware vector representations from the code snippets.

\subsection{Classifiers and CodeBERT Fine-Tuning}
In this subsection, we explain the algorithms used in ML training with software code metrics and the configurations applied to fine-tune the CodeBERT model.
During the development of ML models using software code metrics, we employed SVM(Support Vector Machine), XGBoost, and Passive Aggressive Classifier(PAC) and optimized their parameters through hyperparameter tuning. For SVM (LinearSVC), we applied L2 penalty and squared\_hinge loss function. For XGBoost, we set the objective to "multi:softprob" for multi-class classification and applied tree\_method "hist" to accelerate tree-based learning. We configured the PAC with loss "hinge" and told 1e-3. To ensure the reproducibility of our experiments, we kept the random\_state parameters fixed across all models.

We fine-tuned the codebert-base model from Hugging Face \cite{feng2020codebert}. The training process was conducted over 20 epochs using CrossEntropyLoss, with AdamW as the optimizer and a Linear Learning Rate Decay strategy. In each iteration, the error was minimized through backpropagation.

\subsection{Training Approaches}
In this study, we compared privacy-preserving methods—FL and synthetic data approaches—with centralized learning to investigate whether models trained without data sharing can achieve performance comparable to centrally trained models. 

\textbf{\textit{Centralized Learning with Real Data:}} In this approach, all stakeholders (clients) send their data to a central server, where the received data is aggregated and the model is trained. The trained global model is then distributed back to the clients. This approach enables the development of more generalized models compared to those trained only on client data, as it benefits from a larger dataset. However, it is not applicable in scenarios where data sharing is restricted or sensitive. We performed centralized training to compare its results with FL. Additionally, we also conducted centralized training using only synthetic data.

\begin{figure}[h]
    \centering
    \includegraphics[width=0.8 \textwidth]{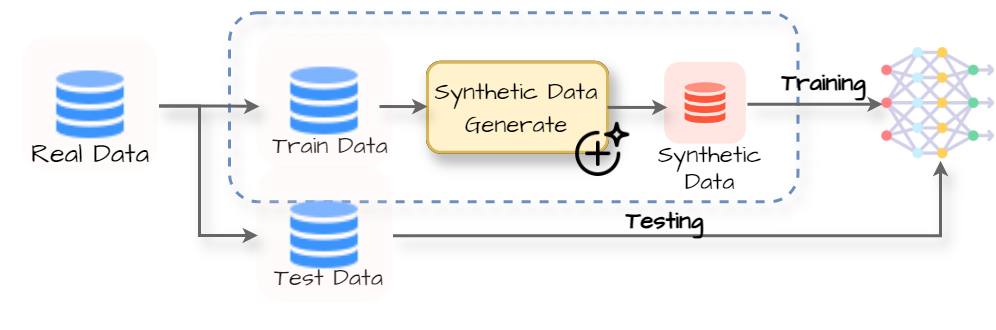}
    \caption{Training with Synthetic Data.}
    \label{fig:arch-synthetic}
\end{figure}

\textbf{\textit{Centralized Learning with Synthetic Data:}}
We aimed to show that training with synthetic data can achieve results comparable to training with real data. To this end, we utilized Gauss Copula, Conditional Tabular Generative Adversarial Network (CTGAN), and CopulaGAN algorithms when applying synthetic data generation to the software code metrics. We then conducted centralized training with the generated synthetic dataset (as illustrated in Figure \ref{fig:arch-synthetic}) and compared the results with those obtained from both centralized and federated training using real data. 

\begin{figure}[h]
    \centering
    \includegraphics[width=0.6 \textwidth]{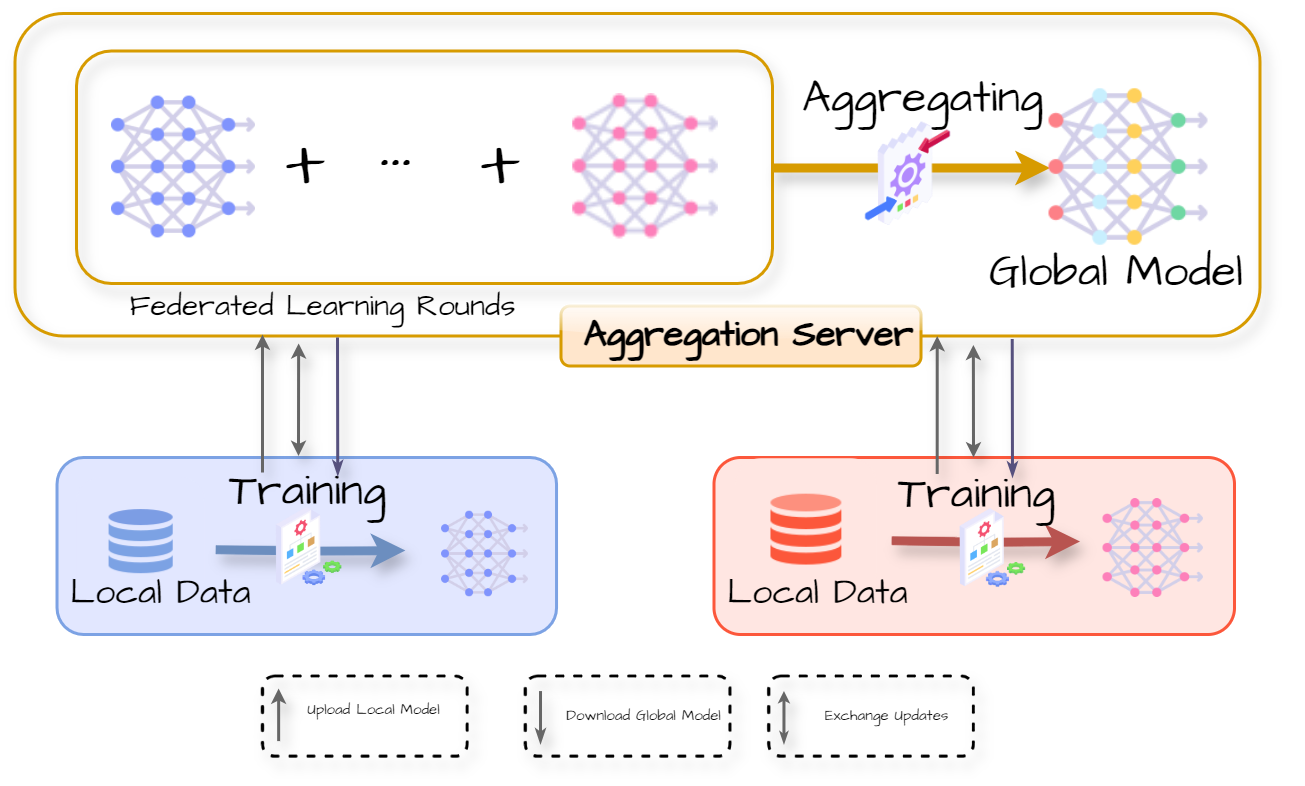}
    \caption{Federated Learning Architecture.}
    \label{fig:arch-FL}
\end{figure}

\textbf{\textit{Federated Learning:}}
Due to data-sharing constraints, local training produces models with low generalization ability and poor performance caused by insufficient data. While centralized training can mitigate this issue, it is often impractical in industrial use cases as it requires data sharing. FL approach enables the development of generalizable models without data sharing while achieving prediction performance comparable to centralized learning. As seen in Figure \ref{fig:arch-FL}, FL process involves clients training models on their local data, sending model parameters to a central server, and the server aggregating these parameters to update and redistribute the global model. This iterative process continues for a predefined number of iterations. Therefore, it enables collaborative model training without data sharing.
We conducted FL with three participants (clients) and partitioned the data set using two different distributions. In the Independent and Identically Distributed (IID) scenario, all participants had a similar number of samples in each class, whereas in the Non-Independent and Non-Identically Distributed (Non-IID) scenario, the number of samples per class varied among participants with Dirichlet allocation (alpha = 1). By implementing these two scenarios, we evaluated the performance of FL under different real-world conditions.

\section{EXPERIMENTAL RESULTS}

In this section, we present our experimental results. We evaluated model performance using the F1-Score (F1), Matthews Correlation Coefficient (MCC), Cohen's Kappa Score, and Geometric Mean (G-Mean). To ensure robustness, we applied 5-fold cross-validation across all training processes and assessed classification performance on the test dataset separated at the beginning of each fold. The reported results in the tables represent the average values obtained over five folds.

\begin{table*}[h]
    \centering
     \caption{Comparison of different training approaches for software code metrics}
    \renewcommand{\arraystretch}{1.3} 
    \setlength{\tabcolsep}{6pt} 
    \resizebox{\textwidth}{!}{ 
    
    \begin{tabular}{lcccccccccccc}
        \toprule
        & \multicolumn{4}{c}{\textbf{Centralized Learning with Real Data}} & \multicolumn{4}{c}{\textbf{Federated Learning}} & \multicolumn{4}{c}{\textbf{Centralized Learning with Synthetic Data}} \\
        \cmidrule(lr){2-5} \cmidrule(lr){6-9} \cmidrule(lr){10-13}
        \textbf{Model} & \textbf{F1} & \textbf{MCC} & \textbf{Kappa Score} & \textbf{G-Mean} & 
        \textbf{F1} & \textbf{MCC} & \textbf{Kappa Score} & \textbf{G-Mean} & 
        \textbf{F1} & \textbf{MCC} & \textbf{Kappa Score} & \textbf{G-Mean} \\
        \midrule
        Linear SVC & 0.47 & 0.10 & 0.09 & 0.54 & 0.46 & 0.07 & 0.06 & 0.53 & 0.45 & 0.04 & 0.04 & 0.52 \\
        Passive Aggressive Classifier & 0.40 & 0.04 & 0.03 & 0.50 & 0.44 & 0.06 & 0.06 & 0.52 & 0.44 & 0.04 & 0.04 & 0.51 \\
        XGBoost & 0.55 & 0.20 & 0.18 & 0.58 & 0.50 & 0.15 & 0.12 & 0.55 & 0.50 & 0.15 & 0.05 & 0.53 \\
        Baseline Model & 0.54 & 0.21 & - & - & - & - & - & - & - & - & - & - \\
        \bottomrule
    \end{tabular}
    }
    \label{tab:comparison}
\end{table*}

Table \ref{tab:comparison} presents a comparative analysis of three different training approaches —Centralized learning with real data, FL, and Centralized Learning with Synthetic Data— applied to software code metrics. Centralized learning with real data approach achieves the highest scores across all models. Nevertheless, FL and centralized learning with synthetic data approaches exhibit competitive results with only minor performance degradation. According to our findings, although centralized learning with real data remains the most effective approach, FL and centralized learning with synthetic data provide promising alternatives for scenarios with data privacy and security issues. Moreover, XGBoost, centralized learning with real data provided the highest performance, with an F1 score of 0.55, MCC of 0.20, and G-Mean of 0.58. It also outperformed the baseline model, which showed similar F1 and MCC results

The results presented in Table \ref{tab:comparison_llm} demonstrate the performance comparison between baseline model, centralized learning, and FL using the CodeBERT model. The baseline model achieved an F1 score of 0.70 and an MCC of 0.51, though it lacks comprehensive evaluation metrics such as Kappa Score and G-Mean. On the other hand, centralized learning outperformed the other approaches, with an F1 score of 0.73, MCC of 0.55, Kappa Score of 0.53, and G-Mean of 0.74. Also, FL, which preserves data privacy by training across distributed nodes, showed slightly lower performance than our centralized learning model and achieved an F1 score of 0.72, MCC of 0.52 Kappa Score of 0.52, and G-Mean of 0.74.

\begin{tcolorbox}[colback=gray!10!white, colframe=black, sharp corners=south, boxrule=0.5mm, width=\linewidth, halign=justify, enlarge left by=0mm, enlarge right by=0mm]
RQ1 Summary: Privacy-preserving methods can achieve results comparable to centralized training for predicting bug severity without sharing data. While centralized learning with real data performed best, privacy-preserving methods performed similarly, with only minor differences.
\end{tcolorbox}

\begin{table}[h]
    \centering
    \renewcommand{\arraystretch}{1.3}
    \caption{CodeBERT Comparison Results}
    \setlength{\tabcolsep}{6pt} 
    \begin{tabular}{p{1.6cm}cccc}
        & \textbf{F1} & \textbf{MCC} & \textbf{Kappa Score} & \textbf{G-Mean} \\
        \midrule
        \textbf{Baseline Model} & 0.70 & 0.51 & - & - \\
        \textbf{Centralized Learning} & 0.73 & 0.55 & 0.53 & 0.74 \\
        \textbf{Federated Learning} & 0.72 & 0.52 & 0.52 & 0.74 \\
        \bottomrule
    \end{tabular}
    \label{tab:comparison_llm}
\end{table}

\section{CONCLUSION and FUTURE WORKS}
In our study, we proposed using FL and synthetic data to develop AI-powered models while preserving data privacy in restricted data-sharing scenarios. Our experiments with ML algorithms and LLMs showed that FL can achieve success without data sharing. We also demonstrated that synthetic data training achieves performance comparable to centralized training. Our findings confirm that LLMs perform best for bug severity prediction, and FL can reach similar scores. Additionally, training with synthetic data ensures privacy while maintaining strong performance.

In future works, we aim to enhance model generalization performance using more advanced synthetic data generation methods. We also plan to explore different aggregation strategies for FL and analyze the applicability of synthetic data and FL methods in solving various challenges in software development processes.

\bibliographystyle{unsrt}  
\bibliography{references}

\end{document}